\documentclass[12pt]{article}
\newtheorem{theorem}{Theorem}

\let\at=@
\usepackage{amsfonts}
\usepackage{latexsym}
\usepackage{amstex}
\newcommand{\dual}[1]{\overset{*}{#1}{}}
\newcommand{\antiself}[1]{\overset{-}{#1}{}}
\newcommand{\self}[1]{\overset{+}{#1}{}}
\newcommand{\uone}[1]{\underset{1}{#1}{}}
\newcommand{\utwo}[1]{\underset{2}{#1}{}}
\newcommand{\sfrac}[2]{\textstyle\frac{#1}{#2}}
\newcommand{\C}{\mathbb C}
\newcommand{\R}{\mathbb R}
\newcommand{\weyl}{\mathfrak W}
\newcommand{\ric}{\mathfrak R}
\renewcommand{\Im}{\operatorname{Im}}
\begin{document}
\title{GEOMETRIC INTERPRETATION OF THE MIXED INVARIANTS OF THE RIEMANN SPINOR}
\author{
Barry M Haddow\thanks{{\sl Email:}haddow\at maths.tcd.ie}
\\School of Mathematics \\
Trinity College\\ Dublin 2\\ IRELAND
}
\maketitle
\begin{abstract}
Mixed invariants are used to classify the Riemann spinor in the
case of Einstein-Maxwell fields and perfect fluids. In the Einstein-Maxwell
case these mixed invariants provide information as to the relative
orientation of the gravitational and electromagnetic principal
null directions. Consideration of the perfect fluid case leads to some results
about the behaviour of the Bel-Robinson tensor regarded as a quartic form
on unit timelike vectors.

\end{abstract}
\thispagestyle{empty}
\newpage
\setcounter{page}{1}
\section{Introduction}

Suppose that $(M,g)$ is a spacetime admitting a spin structure
and denote the Weyl  and Ricci spinors by $\Psi_{ABCD}$ and
$\Phi_{ABC'D'}$ respectively (the abstract index
convention will be used throughout). The most common type of
 {\sl mixed invariant} of the Riemann spinor
is, roughly speaking, a real or complex scalar constructed
from contractions of the Weyl spinor, its complex conjugate
and the Ricci spinor, for example the expression
$\Psi_{ABCD}\Phi^{AB}{}_{A'B'}\Phi^{CDA'B'}$ defines a mixed invariant.
In section two a  precise definition of a mixed invariant
will be given and their significance will be discussed. The main
aim of this paper is to use mixed invariants to
 provide a classification of the Riemann
spinor in the case of spacetimes representing Einstein-Maxwell  fields and
perfect fluids. This classification will be presented in a fashion
which is essentially algorithmic and geometric in character and
will provide a geometric interpretation for the mixed invariants.
Various aspects of such mixed invariants have been considered by
other authors and most recently by Carminati and McLenaghan~\cite{c.mcl};
McIntosh and Zakhary~\cite{mci.z}
and 	Harvey~\cite{harvey} (see also~\cite{penrose}).

Since this paper is only concerned with algebraic properties
of the spinors (and tensors) involved it may be assumed that one is
working at some particular point of the spacetime manifold.
The algebraic classification of the Ricci and Weyl spinors is well
known and has been in  use in relativity for many years. For details
and references see \cite{p.r2,kramer} (for the Petrov classification
of the Weyl spinor/tensor) and \cite{hall.ricci} (for a review
of various methods for classifying the Ricci spinor/tensor).
In addition it is possible to effect these classifications
in a more or less algorithmic fashion~\cite{din.rc,j.mac,let.mcl}
using {\sl pure invariants} of the Weyl and Ricci spinors. A pure invariant
of the Weyl tensor will be defined in section two but
the main ones are, roughly speaking,
 scalars constructed by contraction of copies of
the Weyl spinor, e.g. $\Psi_{ABCD}\Psi^{ABCD}$, and
 pure invariants of the Ricci spinor can be defined analogously.
All these classification schemes must be invariant under $SL(2,\C)$
otherwise the schemes would change appearance under a change of dyad.

Despite the large body of literature concerned with the
classification of the Ricci and Weyl spinors individually, there
has been little or no effort directed to the classification problem
for the full Riemann spinor. The complex vector space of Riemann spinor
type objects splits as a sum of four irreducible subspaces under the action
of $SL(2,\C)$~\cite{p.r} and the components of this decomposition correspond to
the Weyl spinor, its complex conjugate, the Ricci spinor and Ricci scalar.
 Classification
of the Riemann spinor involves the algebraic classification of
its irreducible parts, followed by some consideration
of the `angle' between the parts. By this it is meant that the
Ricci and Weyl spinors each  determine certain directions in spin space
 and a full classification of the Riemann spinor must
take into account the degree of alignment between these directions.
Since the Ricci scalar
is just a real number and contains no geometrical information it will
take no further part in these discussions.
Unfortunately different Weyl and Ricci types require separate treatment
in the consideration of `alignment' and this leads to rather a lot of cases
to be considered. To reduce the amount of work, only three Ricci types
will be studied -  those representing null Einstein-Maxwell fields,
 non-null Einstein-Maxwell fields
and perfect fluids. If these three Ricci types are combined with the five
possible Petrov types
then one arrives at a total of fifteen cases to be considered.
These cases are covered in sections 4,5 and 6 where the NPspinor
computer algebra
package~\cite{npspinor} has been used for some of the calculations.
It should be noted that all the subcases discussed in these sections may not be
realisable in actual spacetimes since restrictions are placed by the
Bianchi identities.

As a way of a gentle introduction to the full classification problem,
a `toy problem' will be considered in section three.
 This toy problem has the benefit of being
of some interest in its own right, as well as illustrating some of the
important aspects of the Riemann spinor classification.  The problem
considered in section three is that of the algebraic classification
of an ordered pair of symmetric 2-spinors $(\sigma_{AB}, \rho_{AB})$.
Since such a pair corresponds to a complex bivector, this
problem is relevant to complex general relativity and, in fact, has
been studied using other methods by Hall~\cite{hall.cx}.
The classification scheme presented in the third section will involve
classifying the individual 2-spinors separately, and then introducing
a concept of angle between the pair of 2-spinors.

\section{Mixed Invariants}

The concept of a `curvature invariant' can be taken to mean
a (generally real or complex) scalar constructed from the curvature
components which is invariant under the action of an appropriate
group of transformations. Such curvature invariants may be appropriate
for singularity theory in that singular points may be detected by
`bad' behaviour of these invariants. Other types of curvature invariants
may be required in renormalisation theory~\cite{fulling}.
The approach taken in this work has more in common with the discussion
of invariants in the context of algebraic classification given by Penrose
and Rindler~\cite{p.r2}. The concept of an algebraic invariant discussed
in this paper should not be confused with the (related) classical
concept of a `differential concomitant' (see~\cite{masque} for
a modern treatment). It is worth reiterating that throughout this paper
it is assumed that one is working at a particular point in the
spacetime manifold.

The  vector space  of  unprimed totally symmetric 4-spinors
(i.e. Weyl-type spinors) will be denoted by $\weyl$ and the
vector space of spinors with 2 primed and 2 unprimed indices,
which are totally symmetric over both primed and unprimed indices,
will be denoted by $\ric$ (corresponding to Ricci-type spinors).
The complex conjugate of $\weyl$ will simply be denoted by $\weyl'$.
A complex
{\sl pure invariant} of the Weyl spinor (of order $p$) is a complex valued
multilinear form on $\weyl$ which is  of $p^{\rm th}$ order in the
Weyl spinor components, and is invariant under the action of
$SL(2,\C)$ on $\weyl$. A real pure invariant of
the Weyl spinor can be defined in a similar fashion.
 The most important pure invariants of the Weyl spinor
are traditionally denoted by $I$ and $J$ and are defined by
\begin{equation}
\label{eq.weylpure}
I=\Psi_{ABCD}
\Psi^{ABCD}\qquad J=\Psi_{AB}{}^{CD}\Psi_{CD}{}^{EF}\Psi_{EF}{}^{AB}
\end{equation}
It is remarked that a pure Weyl invariant, as defined
above,  need not necessarily be  defined by contraction of the Weyl spinor
with copies itself, but pure Weyl invariants defined in such a fashion are
the only ones required in this work.
A pure invariant of the Ricci spinor of order $p$ can be defined in an
analogous fashion
and  the only one required will be the second order real invariant
$R_1$ defined by
\begin{equation}
\label{eq.ricpure}
R_1=\Phi_{ABA'B'}\Phi^{ABA'B'}
\end{equation}

A complex {\sl basic mixed invariant} of order $(p,p',q)$ is an $SL(2,\C)$
invariant multilinear function $M$ defined by
\begin{equation}
\label{eq.mixeddef}
M:\underbrace{\weyl\times\cdots\times\weyl}_{p\ \rm times}\times
\underbrace{\weyl'\times\cdots\times\weyl'}_{p'\ \rm times}\times
\underbrace{\ric\times\cdots\times\ric}_{q\ \rm times}\mapsto\C
\end{equation}
where it will be assumed that $(p+p')q\neq 0$. A real basic
mixed invariant can be defined similarly. The appropriate basic
mixed invariants which will be required consist of the complex functions
\begin{eqnarray}
M_1&=&\Psi_{ABCD}\Phi^{AB}{}_{A'B'}\Phi^{CDA'B'}\label{eq.mixed1}\\
M_2&=&\Psi_{ABEF}
\Psi^{EF}{}_{CD}\Phi^{AB}{}_{A'B'}\Phi^{CDA'B'}\label{eq.mixed2}\\
M_5&=&\Psi_{ABEF}\Psi^{EF}{}_{CD}\Phi^{AB}{}_{A'B'}\Phi^{CD}{}_{C'D'}
\bar{\Psi}^{A'B'C'D'}\label{eq.mixed5}
\end{eqnarray}
and the real functions
\begin{eqnarray}
M_3&=&\Psi_{ABCD}
\bar\Psi_{A'B'C'D'}\Phi^{ABA'B'}\Phi^{CDC'D'}\label{eq.mixed3}\\
M_4&=&\Psi_{ABEF}\Psi^{EF}{}_{CD}\bar\Psi_{A'B'E'F'}\bar\Psi^{E'F'}{}_{C'D'}
\Phi^{ABA'B'}\Phi^{CDC'D'}\label{eq.mixed4}
\end{eqnarray}
The   notation $M_1,\ldots M_5$ is as used in~\cite{c.mcl}
except in that reference the authors used a lower case $M$ and
a different invariant labelled $M_4$. The definition of $M_4$ given here
is of rather high order, but is required in a few cases to extract
certain relevant geometric information. The two real invariants $M_3$ and
$M_4$ can also be expressed in terms of the Bel-Robinson tensor
$T_{abcd}$~\cite{p.r} and the  Ricci tensor $R_{ab}$
according to
\begin{eqnarray}
M_3&=&\sfrac{1}{4}T_{abcd}R^{ab}R^{cd}\label{eq.mixed3t}\\
M_4&=&\sfrac{1}{4}T_{abef}T^{ef}{}_{cd}R^{ab}R^{cd}\label{eq.mixed4t}
\end{eqnarray}
All the pure invariants and basic mixed invariants may
be calculated,
in a tensor formalism, using the GRTensor package~\cite{grtensor}.

Finally, a real or complex {\sl mixed invariant} is defined as a rational
function constructed from pure invariants and basic mixed invariants. It
is thus an $SL(2,\C)$ invariant rational function of the components of
the Weyl spinor, its conjugate and the Ricci spinor.

To close this section, some general remarks concerning
the use of invariants, within the context of classification
up to some invariance group, will be made. Suppose that a group $G$
acts on a set $S$ and hence decomposes $S$ as a disjoint union of
orbits. If $F$ is a $G$-invariant function  (taking values
in some convenient set - say $\R$) then the value
of $F$ is constant on each orbit. A collection of $G$-invariant  functions
$F_1,\ldots, F_p$ is known as a {\sl complete set of invariants} if
$F_1(x)=F_1(y),\ldots, F_p(x)=F_p(y)$ imply that $x$ and $y$ lie
on the same orbit. This list of $G$-invariant functions can therefore be used
to place each element of  $S$ into its appropriate orbit under $G$.
As a straightforward example, consider the tangent space $T_pM$
under the action of the full Lorentz group ${\mathcal L}$. If the
zero vector is excluded then the real-valued function $f(v)=g(v,v)$ provides
a complete set of invariants for the tangent space under ${\mathcal L}$.
This is because any two vectors of the same size lie
in the same ${\mathcal L}$-orbit, and the function
$f$ is obviously Lorentz invariant. In this particular case one normally
effects a broader classification based on the sign ($+,-$ or $0$) of $f$
but the invariant $f$ provides the `finest possible' classification
of non-zero vectors in $T_pM$ under ${\mathcal L}$.
Another method of classifying a non-zero tangent vector $v$
is to give its nature
(that is: timelike, spacelike or null) together with
$\vert g(v,v)\vert$ if $v$ is non-null.
It is remarked
that there is no obvious real-valued Lorentz invariant function on $T_pM$ which
separates the orbit consisting of null vectors, from the orbit
consisting of the zero vector. Instead, one could use the function
$g$ which takes the value $1$ if the sum of the squares of the components
of a vector is non-zero and $0$ if this sum vanishes. This function is
a perfectly acceptable Lorentz invariant function taking values in
the set $\{0,1\}$. Problems analogous to this problem concerning
the separation of zero and null
vectors also arise in the algebraic classification of the Weyl, Ricci
and Riemann spinors using invariants - for example  in the separation of
Petrov types $N$ and $III$ in the algebraic classification
of the Weyl tensor~\cite{din.rc}.

\section{Classification of a Pair of Symmetric 2-Spinors}

For the purposes of this section suppose that $(\sigma_{AB},\rho_{AB})$
is an ordered pair of symmetric 2-spinors which will both
be assumed non-zero. As was mentioned earlier, this ordered
pair can be identified with a complex bivector $F_{ab}$ by
\begin{equation}
\label{eq.cxbivector}
F_{ab}=\sigma_{AB}\epsilon_{A'B'}+\bar\rho_{A'B'}\epsilon_{AB}
\end{equation}
 The classification of a single (non-zero) symmetric 2-spinor
is well-known and can be effected by consideration of
the pure invariant $I_1=\sigma_{AB}\sigma^{AB}$. If $I_1=0$
then $\sigma_{AB}$ is said to be {\sl null} and can be
written as $\sigma_{AB}=\alpha_A\alpha_B$ for some 1-spinor $\alpha_A$
(the repeated principal spinor of $\sigma_{AB}$)
which is defined up to sign. Since all 1-spinors are equivalent under
$SL(2,\C)$ it follows that all null symmetric  2-spinors
are also equivalent under $SL(2,\C)$. If $I_1\neq 0$ then $\sigma_{AB}$
is referred to as {\sl non-null}  can be reduced
to the canonical form $\mu o_{(A}\iota_{B)}$ where $o_A,\iota_A$
is a normalised spin dyad (that is $o_A\iota^A=1$) and $\mu\in\C$.
The dyad is defined up to transformations of the form
$o_A\mapsto\lambda o_A$, $\iota_A\mapsto\lambda^{-1}\iota_A$
for $\lambda\in\C$ and the 1-spinors $o_A$ and $\iota_A$ are principal
spinors of $\sigma_{AB}$.
Such a 2-spinor is therefore uniquely
determined, up to $SL(2,\C)$ transformations, by the complex scalar $\mu$
- given by $-\mu^2=2I_1$. The invariant $I_1$
does not form a complete set since it takes equal values for
$\sigma_{AB}$ and $-\sigma_{AB}$, which lie on different
$SL(2,\C)$ orbits. It is possible to construct some kind of invariant
taking values in $\{-1,1\}$ to separate these two cases but generally such
sign problems may be ignored.

Now turn the case of a pair of symmetric 2-spinors
$(\sigma_{AB},\rho_{AB})$ and define three invariants
in the following manner. Pure invariants $I_1$ and $I_2$ are defined by
\begin{equation}
\label{eq.2spinpure}
I_1=\sigma_{AB}\sigma^{AB}\qquad I_2=\rho_{AB}\rho^{AB}
\end{equation}
and a mixed invariant $M$ is defined by
\begin{equation}
\label{eq.2spinmixed}
M=\sigma_{AB}\rho^{AB}
\end{equation}
In general  the set $\{I_1,I_2,M\}$ is complete except for the sort of
sign problems which arose in the case of a single symmetric 2-spinor.
The cases where this set is not complete are of interest since similar
problems arise in the case of the Riemann spinor. For the
pair $(\sigma_{AB},\rho_{AB})$ there are really four cases to consider, denoted
in an obvious fashion as non-null/non-null, null/non-null, non-null/null
and null/null. The middle two cases in this list can clearly be dealt
with in the same fashion so one is left with three cases to consider.

\paragraph*{Case 1: Non-null/Non-null}

In this case $I_1\neq 0$ and $I_2\neq 0$ and it may be assumed that there
exist 1-spinors $\alpha_A,\beta_A,\gamma_A$ and $\delta_A$ such that
\begin{equation}
\label{eq.nonnon}
\sigma_{AB}=\alpha_{(A}\beta_{B)}\qquad\rho_{AB}=\gamma_{(A}\delta_{B)}
\end{equation}
Given four ordered
 spinor directions, one can define their {\sl cross-ratio}~\cite{p.r,semple}
in terms of 1-spinors spanning these directions. The cross-ratio
of the four directions spanned by $\alpha_A,\beta_A,\gamma_A$ and $\delta_A$
is denoted by $\chi$ and defined by
\begin{equation}
\label{eq.cross}
\chi=\frac{(\alpha_A\gamma^A)(\beta_B\delta^B)}
{(\alpha_C\delta^C)(\beta_D\gamma^D)}
\end{equation}
The function $\chi$ is assumed to take values in $\C\cup\{\infty\}$.
If the corresponding flag-pole directions are denoted by $(A,B,C,D)$
then the cross-ratio can be interpreted geometrically in terms of
Lorentz transformations  required to take the $A-B$ plane into the
$C-D$ plane~\cite{p.r}. The cross-ratio $\chi$ of
$\alpha_A,\beta_A,\gamma_A$ and $\delta_A$ can
be expressed in terms of $I_1,I_2$ and $M$ by
\begin{equation}
\label{eq.2spincross}
\chi=\frac{\mu+1}{\mu-1}
\qquad\quad{\rm where}\quad \mu^2=\frac{\hfill M^2}{I_1I_2}
\end{equation}
The sign ambiguity in the above definition of $\chi$ corresponds
to the arbitrary ordering of $\alpha_A$ with respect to $\beta_A$ and
of $\gamma_A$ with respect to $\delta_A$. If the directions
of two or more of the 1-spinors $\alpha_A,\beta_A,\gamma_A$ and $\delta_A$
coincide then $\chi$ must take one of the values
$0$, $1$ or $\infty$. Since one cannot have $\chi=1$ for finite
$\mu$ these degenerate cases correspond to $\mu=\pm 1$ or equivalently
$M^2=I_1I_2$. One thus concludes that {\sl the condition $M^2=I_1I_2$
is necessary and sufficient for at least one of the
principal spinors of $\sigma_{AB}$ to align with one of the principal
spinors of $\rho_{AB}$}.

The case where $M^2\neq I_1I_2$ is referred to as the {\sl non-aligned}
case and in this case the pair of symmetric 2-spinors
is determined (up to sign)  by the set of three complex invariants
$\{I_1,I_2,M\}$. The expression (\ref{eq.2spincross}) determines the
cross-ratio
and hence the relative positions of the directions of the four principal
spinors
up to $SL(2,\C)$ rotations~\cite{p.r} and then the invariants $I_1$ and $I_2$
determine individual scalings.

The aligned
($M^2=I_1I_2$) case can be further subdivided into the {\sl partially
aligned} case where exactly one of the principal spinors of $\sigma_{AB}$
is parallel to exactly one of the principal spinors of $\rho_{AB}$; and
the {\sl totally aligned} case where $\sigma_{AB}$ is a multiple of
$\rho_{AB}$.
In the partially aligned case, if one notes that any three spinor directions
are equivalent (under $SL(2,\C)$) to any other then it is clear that
knowledge of $I_1$ and $I_2$ is sufficient to reconstruct the spinor pair
(possibly with some residual sign ambiguity).
Similar comments apply in the totally aligned case. The problem is
that the invariants $I_1,I_2$ and $M$ do not appear to be able
to separate the partially
and totally aligned cases from each other. One method of
resolving this problem is to consider the symmetric 2-spinor $\phi_{AB}$
defined by $\phi_{AB}=\sigma_{AC}\rho^C{}_{B}-\rho_{AC}\sigma^C{}_B$
and define an invariant $s$ which takes the value $1$ if
$\phi_{AB}\neq 0$ and $0$ if $\phi_{AB}=0$. It may be verified
that $s=0$ if and only if $\sigma_{AB}$ is a multiple
of  $\rho_{AB}$.

then show that $\phi_{AB}=(\beta_C\gamma^C)\alpha_A\alpha_B$

\paragraph*{Case 2: Null/Non-null}

In this case one has that $I_1=0$ and $I_2\neq 0$ and one can choose a
dyad $o_A$, $\iota_A$ (with $o_A\iota^A=1$) such that
\begin{equation}
\label{eq.nonnul}
\sigma_{AB}=(o_A+B\iota_A)(o_B+B\iota_B)\qquad\rho_{AB}=\mu o_{(A}\iota_{B)}
\end{equation}
for $B,\mu\in\C$. The pair of 2-spinors are then completely determined,
up to $SL(2,\C)$ transformations, by the complex quantities $B$ and $\mu$
which are related to the invariants $M$ and $I_2$ by $M=-\mu B$ and
$I_2=-\mu^2/2$. Thus $M$ and $I_2$ form a complete set of
invariants (apart from a sign ambiguity) and furthermore there is an
{\sl aligned} case corresponding to $M=0$.

\begin{table}
\begin{tabular}{ l c c c c}         \hline
 Type & $I_1$ & $I_2$ & $M$ & $\phi_{AB}$
 \\    \hline \hline
 Non-null/Non-null aligned & $\neq 0$ & $\neq 0$ & $\neq I_1I_2$ & \\ \hline
 Non-null/Non-null
 partially aligned & $\neq 0$ & $\neq 0$ & $=I_1I_2$ & $\neq 0$
\\ \hline
 Non-null/Non-null totally aligned & $\neq 0$ & $\neq 0$ & $=I_1I_2$ & $=0$
\\ \hline
 Null/Non-null non-aligned & $=0$ & $\neq 0$ & $\neq 0$ & \\ \hline
 Null/Non-null aligned & $=0$ & $\neq 0$ &  $=0$ & \\ \hline
 Null/Null non-aligned & $=0$ & $=0$ & $\neq 0$ & \\ \hline
 Null/Null aligned & $=0$ & $=0$ & $=0$ & \\ \hline
 \end{tabular}
\caption{Classification of a pair of 2-spinors. The invariants $I_1, I_2$
and $M$ are defined by equations (\ref{eq.2spinpure}) and (\ref{eq.2spinmixed})
and $\phi_{AB}\equiv\sigma_{AC}\rho^C{}_B-\rho_{AC}\sigma^C{}_{B}$}
\label{twospinor}
\end{table}

\paragraph*{Case 3: Null/Null}

The final case is characterised by $I_1=I_2=0$ and the pair of
2-spinors take the form
\begin{equation}
\label{eq.nulnul}
\sigma_{AB}=\alpha_A\alpha_A\qquad\rho_{AB}=\beta_A\beta_B
\end{equation}
This case can be split into the {\sl aligned case} ($\alpha_A$ parallel
to $\beta_A$, equivalent to $M=0$) and the {\sl non-aligned case} ($M\neq 0$).
In the non-aligned case, the 2-spinor pair is determined by the
pair of 1-spinors $\alpha_A$ and $\beta_A$. However, a pair of non-parallel
1-spinors are determined uniquely (up to $SL(2,\C)$ transformations)
 by their inner product and
this is given by $M=(\alpha_A\beta^A)^2$. There is a sign ambiguity in
the inner product, but the individual signs of $\alpha_A$ and $\beta_A$
are irrelevant in the determination of $\sigma_{AB}$ and $\rho_{AB}$.
In the non-aligned case,
the mixed invariant $M$ is therefore sufficient to determine
the pair of 2-spinors up to $SL(2,\C)$ transformations.
In the aligned case, one has $\rho_{AB}=\lambda\alpha_A\alpha_B$ for some
$\lambda\in\C$.
There are clearly no non-zero invariants which can be constructed
by contractions of the pair of 2-spinors but  different values
of $\lambda$ give rise to inequivalent (under $SL(2,\C)$) pairs
of symmetric 2-spinors. With respect to any given dyad the components
of $\rho_{AB}$ must be $\lambda$ times the components of $\sigma_{AB}$.
Hence one can calculate $\lambda$ by choosing ${\bf A}, {\bf B}$
such that $\rho_{\bf AB}\neq 0$ (bold-face indices
denote components) and then $\lambda=\rho_{\bf AB}/\sigma_{\bf AB}$.
The quantity $\lambda$ defined in this way is independent of
${\bf A}$ and ${\bf B}$ and of the dyad used and
$\{\lambda\}$ is a complete set of invariants in the aligned case.

A summary of the discussion in this section is given in table \ref{twospinor}
which displays the classification scheme for a pair of
non-zero symmetric 2-spinors using the invariants $I_1,I_2$ and $M$.

\section{Null Einstein-Maxwell Fields}

Throughout this section it will be assumed that the Ricci spinor takes
the form $\Phi_{ABC'D'}=\gamma_A\gamma_B\bar\gamma_{C'}\bar\gamma_{D'}$
for some 1-spinor $\gamma_A$. Clearly the Ricci spinor is completely
determined by $\gamma_A$ and in fact $\lambda\gamma_A$ will give rise
to the same Ricci spinor provided that $\vert\lambda\vert=1$.
Given that the Ricci spinor is of this form, the necessary mixed
invariants required to analyse the relationship between the Weyl and Ricci
spinors turn out to be $M_3$, $M_4$ and $M_5$. The expressions
for these invariants are
\begin{eqnarray}
M_3&=&\left\vert\Psi_{ABCD}\gamma^A\gamma^B\gamma^C\gamma^D\right\vert^2
\label{eq.nullm3}\\
M_4&=&
\left\vert\Psi_{ABEF}\Psi^{EF}{}_{CD}
\gamma^A\gamma^B\gamma^C\gamma^D\right\vert^2
\label{eq.nullm4}\\
M_5&=&\left(\Psi_{ABEF}\Psi^{EF}{}_{CD}\gamma^A\gamma^B\gamma^C\gamma^D\right)
\left(\bar\Psi_{A'B'C'D'}\bar\gamma^{A'}\bar\gamma^{B'}\bar\gamma^{C'}
\bar\gamma^{D'}\right)\label{eq.nullm5}
\end{eqnarray}
In the above one has that $M_3M_4=\vert M_5\vert^2$ and the invariants
$M_3$ and $M_5$ (perhaps together with some pure Weyl invariants)
will be seen to
form a complete set of invariants for the Riemann spinor in the
case of a null Einstein Maxwell field (cf~\cite{c.mcl}). The invariants
$M_3$ and $M_4$ have a direct geometric interpretation, as will
be seen in theorem~\ref{th.nullem}.

If $\gamma_A$ is a repeated
principal spinor of the Weyl spinor then the Ricci and Weyl spinors will
be said to be {\sl repeatedly aligned} and similarly the terminology
{\sl non-repeatedly aligned} will refer to the case where $\gamma_A$
is a non-repeated principal spinor of the Weyl spinor. It should
be noted that Maxwell's equations imply that the flagpole direction
of $\gamma_A$ is geodesic and shearfree and hence is a repeated
principal
null direction of the Weyl tensor (by the Goldberg-Sachs theorem~\cite{g.s})
 and one is forced
into the repeatedly aligned case. In the case of a null fluid
one does not have Maxwell's equations and all
cases - repeatedly aligned, non-repeatedly aligned and
non-aligned may be possible. The following theorem
shows how $M_3$ and $M_4$ may be used to separate these three possibilities.

\begin{theorem}
\label{th.nullem}
Suppose that $\Phi_{ABC'D'}=\gamma_A\gamma_B\bar\gamma_{C'}\bar\gamma_{D'}$
and that $M_3$ and $M_4$ are as defined by equations (\ref{eq.nullm3})
and (\ref{eq.nullm4}). One has the following three possibilities
for alignment between the Weyl and Ricci spinors, assuming both
to be non-zero.
\begin{description}
\item[{\it (i)  Non-aligned}] $M_3\neq 0$, $M_4\neq 0$
\item[{\it (ii)  Non-repeatedly aligned}] $M_3=0$, $M_4\neq 0$
\item[{\it (ii)  Repeatedly aligned}] $M_3=M_4=0$
\end{description}
\end{theorem}
\paragraph*{Proof}

The condition $M_3=0$ is clearly equivalent to
(repeated or non-repeated) alignment. If $M_3=0$ then let
$\eta_{AB}=\Psi_{ABCD}\gamma^C\gamma^D$ and it follows that
$\eta_{AB}\gamma^A\gamma^B=0$ and
$\eta_{AB}\eta^{AB}=0\Leftrightarrow M_4=0$. But then $M_4=0$
implies  $\eta_{AB}\propto\gamma_A\gamma_B$  and hence $\eta_{AB}\gamma^A=0$.
The definition of $\eta_{AB}$ then shows that $\Psi_{ABCD}\gamma^A
\gamma^B\gamma^C=0$ i.e. $\gamma^A$ is a repeated principal spinor
of $\Psi_{ABCD}$. Conversely if $\gamma_A$ is a repeated principal
spinor of $\Psi_{ABCD}$ it follows that $\eta_{AB}\gamma^A=0$ and
hence $\eta_{AB}$ is null and so $M_4=0$. $\Box$

For each of the Petrov types the Weyl spinor invariantly determines
a normalised
dyad $o_A,\iota_A$ up to some, possibly trivial, group of transformations.
The spinor $\gamma_A$ can be written in this dyad as
$\gamma_A=A\iota_A-Bo_A$ so that $A=o_A\gamma^A$ and $B=\iota_A\gamma^A$
and, given the indeterminacy in $\gamma_A$,  $A$ and $B$ are defined up to
transformations of the form $A\rightarrow\lambda A$ and
$B\rightarrow\lambda B$ with $\lambda\in\C$, $\vert\lambda\vert=1$.
For each Petrov type, the appropriate pure Weyl invariants determine
the Weyl spinor and then one merely needs to find $A$ and $B$ in order to
determine the Ricci spinor and hence the full Riemann spinor. In general
$A$ and $B$ can be expressed in terms of the mixed invariants $M_3,M_4$ and
$M_5$ and the following is a case by case account of how this is done.

\paragraph*{Petrov type $N$}

This is arguably the simplest case
 since both the Ricci spinor and Weyl spinors are each
determined by a single 1-spinor. The Weyl spinor can be written
as $\Psi_{ABCD}=o_Ao_Bo_Co_D$ for some 1-spinor $o_A$, defined
up to multiplication by $\lambda\in\C$ satisfying $\lambda^4=1$. A dyad
$o_A,\iota_A$ is determined by the Weyl spinor in which the 1-spinor
$\iota_A$ is arbitrary apart from the normalisation condition
$o_A\iota^A=1$. In the non-aligned case, where $o_A\gamma^A\neq 0$,
the 1-spinor $\iota_A$ may be assumed to be parallel $\gamma_A$,
that is $\gamma_A$ can be written as $\gamma_A=A\iota_A$
for some $A\in\R$, $A>0$. Under these assumptions one has that $M_3=A^8$
 and hence $M_3$ completely determines the relationship between
$\gamma_A$ and $o_A$ and therefore determines the Riemann spinor
up to $SL(2,\C)$ transformations. In the
(necessarily repeatedly) aligned case the Ricci spinor
takes the form $\lambda o_Ao_B\bar{o}_C'\bar{o}_D'$ for some $\lambda\in\R$.
Clearly there are no non-zero mixed invariants which can be formed
from contractions of the Weyl and Ricci spinors in this case, and
yet different values of $\lambda$ give rise to $SL(2,\C)$-inequivalent
Riemann spinors. In the aligned case the invariants formed by
contractions do not form a complete set and this is reminiscent
of the null/null aligned case in section two and
one can use similar methods to define the invariant $\lambda$.

\paragraph*{Petrov type $D$}

In this case the Weyl spinor is determined by a pair of spin directions
and a complex scalar. One can choose a normalised dyad $o_A,\iota_A$
spanning the principal directions of the Weyl spinor and determined
up to the transformations $o_A\rightarrow\lambda o_A$ and
$\iota_A\rightarrow\lambda^{-1}\iota_A$ for $\lambda\in\C$. This
dyad may be assumed scaled  (and reordered if necessary)
so that the 1-spinor $\gamma_A$
can be written as $\gamma_A=A\iota_A-o_A$ where $A$ is real and
non-negative. The Weyl spinor may be assumed to take the form
$6\eta o_{(A}o_B\iota_C\iota_{D)}$ for $\eta\in\C$ and then the pure
Weyl invariants defined by (\ref{eq.weylpure}) are given by
$I=6\eta^2$ and $J=-6\eta^3$. Now
 a short
 calculation will  show that $M_3=6A^4\vert I\vert$ and hence the invariants
$M_3$ and $I$ will determine the Weyl spinor (up to sign) and the Ricci spinor.
The (necessarily repeatedly) aligned case corresponds to $M_3=0$
and in this case the only invariant required to determine the  Ricci and Weyl
spinors (up to sign) is $I$.
If one wishes a complete set of invariants, that is if one wishes
to remove the sign ambiguity, then one can replace $I$ with $J/I$.

\paragraph*{Petrov type $III$}

In the Petrov type $III$ case the Weyl spinor determines (and is determined by)
an ordered normalised dyad $o_A,\iota_A$ subject to the transformation
$o_A\rightarrow -o_A$, $\iota_A\rightarrow -\iota_A$. The canonical
form in this dyad is $\Psi_{ABCD}=-4o_{(A}o_Bo_C\iota_{D)}$
where $o_A$ is a repeated principal spinor and  $\iota_A$ is
a non-repeated principal spinor. Writing $\gamma_A=A\iota_A-Bo_A$ as before one
has
\begin{eqnarray}
M_3&=&\vert 4A^3B\vert^2\label{eq.null3m3}\\
M_4&=&\vert 2A^4\vert^2\label{eq.null3m4}\\
M_5&=&(\overline{4A^3B})(2A^4)\label{eq.null3m5}
\end{eqnarray}
It can be
confirmed from (\ref{eq.null3m4}) that $M_4=0$ is the necessary and sufficient
condition for repeated alignment and that $M_3\neq 0$ is equivalent
to no alignment. In  the non-aligned case
and the non-repeatedly aligned case $M_3$ and $M_5$ determine
$4A^3B$ and $2A^4$ up to multiplication by a complex number of unit modulus.
Consequently $M_3$ and $M_5$ determine $A$ and $B$ to within their intrinsic
ambiguity and so form a complete set in this case. However,
in the repeatedly aligned case  $A=0$ and
then all three of the mixed invariants $M_3$, $M_4$ and $M_5$ vanish
and one has $\Phi_{ABC'D'}=\lambda o_Ao_B\bar{o}_{C'}\bar{o}_{D'}$ for
some $\lambda\in\R$.
It can be seen that there are no mixed non-zero mixed invariants
which can be expressed in terms of contractions of the
Ricci and Weyl spinors and the  invariant function $\lambda$
must be determined using similar methods to those used in the
null/null aligned case in section two.

\paragraph*{Petrov type $II$}

In this case a canonical dyad $o_A,\iota_A$ is determined to within a finite
number of discrete transformations~\cite{p.r2} and in this dyad
the Weyl spinor takes the form
\begin{equation}
\label{eq.null2weyl}
\Psi_{ABCD}=6\eta ( o_{(A}o_B\iota_C\iota_{D)}+o_Ao_Bo_Co_D)
\end{equation}
The repeated principal spinor  in this case is $o_A$; the non-repeated
principal spinors are $o_A\pm i\iota_A$
and the pure Weyl invariants are $I=6\eta^2$ and $J=-6\eta^3$.
If $\gamma_A=A\iota_A-Bo_A$
then the mixed invariants $M_3,M_4$ and
$M_5$ are given by
\begin{eqnarray}
M_3&=&6\vert I\vert \vert P_1\vert^2\label{eq.null2p3}\\
M_4&=&\vert I\vert^2\vert P_2\vert^2\label{eq.null2p4}\\
\bar{I}M_5&=&-6I\bar{J}\bar{P_1}P_2\label{eq.null2p5}
\end{eqnarray}
where
\begin{equation}
\label{eq.null2p}
P_1=A^4+B^2A^2\qquad{\rm and}\qquad P_2=2A^4-B^2A^2
\end{equation}
Assuming that $I$ and $J$ are given, equations
(\ref{eq.null2p3})-(\ref{eq.null2p5}) provide values for the expressions
$\vert P_1\vert^2$, $\vert P_2\vert^2$ and $\bar{P_1}P_2$ and hence determine
$P_1$ and $P_2$ subject to the transformations
$(P_1,P_2)\rightarrow (P_1e^{i\theta},P_2e^{i\theta})$ for
$\theta\in\R$. However this ambiguity merely reflects the intrinsic ambiguity
in $A$ and $B$ and so one can consider the values of the polynomials
$A^4+B^2A^2$
and $2A^4-B^2A^2$ to be completely determined by the mixed invariants.
These polynomials will in turn determine the values of $A$ and $B$ except in
the
case where $A=0$. This exceptional case is the repeatedly aligned case.

In the non-repeatedly aligned case, $A=\pm iB$ and so it may
be assumed that $B$ is purely
imaginary and that $A$ is real and then $M_4=9\vert I\vert^2A^8$ and hence
the components of $\gamma_A$ in the dyad $o_A,\iota_A$
are determined by $I$ and $M_4$.  The case where $M_4=0$ and
$M_3\neq 0$ corresponds to $\gamma_A$ being a principal spinor of
the square of the Weyl spinor. In the repeatedly aligned
case one again finds that all mixed invariants
constructed by contractions of the Ricci and Weyl spinors vanish
and cannot give a complete classification of the Riemann spinor.

\paragraph*{Petrov type $I$}

If the Weyl spinor is of Petrov type $I$ then one
possible way of putting it into canonical for is to choose a
(unique up to certain discrete transformations~\cite{p.r2}) dyad
$o_A,\iota_A$ where $\Psi_{ABCD}$ takes the form
\begin{equation}
\label{eq.Icanonical}
\Psi_{ABCD}=6\eta(\chi o_{(A}\iota_B\iota_C\iota_{D)}+
(\chi+1)o_{(A}o_B\iota_C\iota_{D)}+ o_{(A}o_Bo_C\iota_{D)})
\end{equation}
for $\eta,\chi\in\C$. In this case the principal spinors
are $o_A, \iota_A, o_A+\iota_A$ and $o_A+\chi\iota_A$
and if one defines $P_1$ and $P_2$ by
\begin{eqnarray}
P_1&=&6\eta AB(A+B)(\chi B+A)\label{eq.p1}\\
P_2&=&-\sfrac{3}{2}\eta^2
[3A^4+(4+4\chi)A^3B+(4+2\chi+4\chi^2)A^2B^2\nonumber\\
   & &\qquad\qquad +(4\chi+4\chi^2)B^3A+3\chi^2B^4]\label{eq.p2}
\end{eqnarray}
then the appropriate pure and mixed invariants are given by
\begin{eqnarray}
M_3&=&\vert P_1\vert^2\label{eq.null1m3}\\
M_4&=&\vert P_2\vert^2\label{eq.null1m4}\\
M_5&=&\bar{P_1}P_2\label{eq.null1m5}\\
I&=&6\eta^2(\chi-\chi+1)\label{eq.null1i}\\
J&=&-6\eta^3(\chi+1)(\chi-2)(\chi-\sfrac{1}{2})\label{eq.null1j}
\end{eqnarray}
In the non-aligned case $A,B,\eta$ and $\chi$ are determined by $I,J,M_3$ and
$M_5$ as can be shown using similar arguments to those employed
in the Petrov type $II$ case. In the exceptional case where
$M_3=0$ then $\gamma_A$ is parallel to one of the principal
spinors of $\Psi_{ABCD}$ and one has
either $A=0$, $B=0$, $B=-A$ or $B=-A/\chi$
and a complete set of invariants is $I,J$ and $M_4$. In these
aligned
cases $M_5=0$ and $M_4$ is non-zero and  given by the following expressions
\begin{eqnarray}
A=0,&\qquad&M_4=\left\vert\sfrac{9}{2}\chi^2\eta^2B^4\right\vert^2
\label{eq.nalign1}\\
B=0,&\qquad&M_4=\left\vert\sfrac{9}{2}\eta^2A^4\right\vert^2\label{eq.nalign2}\\
B=-A,&\qquad&M_4=\left\vert\sfrac{9}{2}\eta^2A^4(\chi-1)^2\right\vert^2
\label{eq.nalign3}\\
A=-B\chi,&\qquad&M_4=
\left\vert\sfrac{9}{2}\eta^2\chi^{2}B^4(\chi-1)^2\right\vert^2
\label{eq.nalign4}
\end{eqnarray}

\section{Non-Null Einstein-Maxwell Fields}

For the purposes of this section it will be assumed that the Ricci spinor
can be written as $\Phi_{ABC'D'}=\phi_{AB}\bar\phi_{C'D'}$
for some non-null symmetric 2-spinor $\phi_{AB}$. The Ricci spinor
therefore picks out two directions in spin space
spanned by $\alpha_A$ and $\beta_A$,  the principal spinors of $\phi_{AB}$.
It
may be assumed  that
 $\alpha_A\beta^A=1$ so that $\phi_{AB}=\mu\alpha_{(A}\beta_{B)}$
where the phase of the complex number $\mu$ is arbitrary and so $\mu$ may
be chosen real and positive. Also the unordered pair of
 principal Ricci spinors
$\alpha_A$ and $\beta_A$ are determined subject to the transformations
$\alpha_A
\rightarrow\lambda\alpha_A$ and $\beta_A\rightarrow\lambda^{-1}\beta_A$.
The Ricci spinor in this case is completely determined (up to $SL(2,\C)$
transformations) by the real invariant $R_1$ which is related
to $\mu$ through the equation $4R_1=\vert\mu\vert^4$. The relevant mixed
invariants for this section will be denoted by $N_1$ and $N_2$
and are defined by $N_1=-M_1/R_1$ and $N_2=M_2/R_1$. These invariants
are given by the following equations.
\begin{eqnarray}
N_1&=&2\Psi_{ABCD}\alpha^A\beta^B\alpha^C\beta^D\label{eq.nonn1}\\
N_2&=&-
2\Psi_{ABEF}\Psi^{EF}{}_{CD}\alpha^A\beta^B\alpha^C\beta^D\label{eq.nonn2}
\end{eqnarray}

\paragraph*{Petrov type $N$}
The Weyl spinor can be written as $\Psi_{ABCD}=o_Ao_Bo_Co_D$ and
determines a canonical normalised dyad $o_A,\iota_A$ in which the direction of
$\iota_A$ is arbitrary, and $o_A$ is subject to multiplication
by a
fourth root of unity. It may therefore be assumed that $\beta_A=\iota_A$, by
scaling $\beta_A$ and/or swapping  $\alpha_A$ and $\beta_A$ as necessary.
The
condition $\alpha_A\beta^A=1$ then shows that there exists $A\in\C$ such that
$\alpha_A=o_A+A\iota_A$ where, given the ambiguity in $o_A$, it may be assumed
that $A$ lies in the first quadrant of the complex plane. The mixed invariant
$N_1$ is then equal to $2A^2$ and hence determines $A$ uniquely
and a complete set of invariants for this case is $\{R_1,N_1\}$. The special
case $A=0$, equivalent to $N_1=0$, is the {\sl aligned case}.

\paragraph*{Petrov type $D$}
In this case  the Weyl spinor
and the Ricci spinor each determine a pair of directions
in spin space. Since any four ordered directions in spin space are
determined by their cross ratio $\chi$ it follows that the Riemann spinor
is determined by $\chi,I$ and $R_1$ (up to a sign ambiguity).
In order to compute the
cross ratio, assume that $\Psi_{ABCD}=6\eta o_{(A}o_B\iota_C\iota_{D)}$,
as in the previous section, and suppose that the principal
Ricci spinors $\alpha_A$ and $\beta_A$ are given by
\begin{equation}
\label{eq.nonnull}
\alpha_A=(AB+1)o_A+B\iota_A\qquad\beta_A=Ao_A+\iota_A
\end{equation}
where $A,B\in\C$. The invariant $N_1$ is then given by
\begin{equation}
\label{eq.nnDn1}
N_1=2\eta(6A^2B^2+6AB+1)
\end{equation}
The cross ratio of the four spinors $o_A,\iota_A,\alpha_A,\beta_A$
can then be expressed in terms of $A$ and $B$ as follows
(recall that $\chi$ takes values in $\C\cup\{\infty\}$)
\begin{equation}
\label{eq.nnDcross}
\chi=\frac{AB}{AB+1}
\end{equation}
A short calculation using (\ref{eq.nnDcross}) and
(\ref{eq.nnDn1}) and the
expressions for $I$ and $J$ in the type $D$ case, given
in the last section, shows that
\begin{equation}
\label{eq.nnDcross2}
\chi+\frac{1}{\chi}=\frac{2N_1I-8J}{N_1I+2J}
\end{equation}
Now recall that if any of the four spin directions coincide
then the cross ratio must take one of the values $0,1$ or $\infty$.
{}From (\ref{eq.nnDcross2}) it can be seen that $\chi\neq 1$ for
$J\neq 0$ (and $J=0$ is inconsistent with Petrov
type $D$) and that $\chi=0$ or $\infty$ is equivalent
to $N_1I+2J=0$. {\sl Thus the  aligned case is characterised
by the condition $N_1I+2J=0$}.

In the non-aligned case, the cross ratio is determined by (\ref{eq.nnDcross2})
and the residual ambiguity in the
value of $\chi$ corresponds to the ambiguity in the ordering
within the pairs $o_A,\iota_A$ and $\alpha_A,\beta_A$. The invariants
$I,R_1$ and $N_1$ therefore form a complete set in this case
(apart from a sign ambiguity).

The aligned case can be divided into {\sl partially aligned},
where exactly one of the principal spinors of the Weyl spinor is
parallel to a principal Ricci spinor, and {\sl totally aligned} where
both principal spinors of the Weyl spinor are parallel to the
Ricci principal spinors. Since any pair of spinor directions
is equivalent under $SL(2,\C)$ and similarly for any triple of
spinor directions it follows that for each type of alignment the
appropriate pure invariants form a complete set. This begs the question
of how one
separates the partially and totally aligned cases and it would appear
that there are no suitable real or complex mixed invariants.
It is relevant to compare this problem with a similar difficulty
which arose during the consideration of the non-null/non-null
aligned case in section 3. In the present situation the
simplest solution seems to be to consider the following expression
\begin{equation}
\label{eq.nnDcomm}
\Phi_{ABE'F'}\Phi^{E'F'GH}\Psi_{GHCD}-\Phi_{CDE'F'}\Phi^{E'F'GH}\Psi_{GHAB}
\end{equation}
If one supposes that the Ricci and Weyl spinors are at least partially
aligned then one may assume that $\Psi_{ABCD}=6\eta o_{(A}o_B\iota_C\iota_{D)}$
and $\Phi_{ABC'D'}=\phi_{AB}\bar\phi_{C'D'}$
where $\phi_{AB}=\mu o_{(A}\beta_{B)}$ and $\beta_A=Ao_A+\iota_A$.
In this case the expression (\ref{eq.nnDcomm}) takes the form
\begin{equation}
\label{eq.nnDcomm2}
\frac{\mu^4\eta}{2}\left[(2o_{(C}\iota_{D)}-Ao_Co_D)o_{(A}\beta_{B)}
-(2o_{(A}\iota_{B)}-Ao_Ao_B)o_{(C}\beta_{D)}\right]
\end{equation}
The expression (\ref{eq.nnDcomm2}) clearly vanishes if $\iota_A$
is parallel to $\beta_A$ (and hence $A=0$). Conversely if (\ref{eq.nnDcomm2})
vanishes then contracting with $\iota^Ao^C$ gives
\begin{equation}
\label{eq.nnDcontract}
\frac{3A\mu^4\eta}{4}o_Bo_D=0
\end{equation}
which implies that $A=0$. Consequently, {\sl in the aligned case
the vanishing of (\ref{eq.nnDcomm}) is equivalent to total alignment}.
It is worth remarking that this result also follows if one notes
that the vanishing of (\ref{eq.nnDcomm}) is equivalent to $\phi_{AB}$
being an eigen-2-spinor of $\Psi_{ABCD}$ and refers to the
remarks in~\cite{p.r2} concerning
the  eigen-2-spinor structure of the Weyl spinor.

\paragraph*{Petrov type $III$}

As in section 4, it may be assumed that $o_A$ and $\iota_A$
are principal spinors and the Weyl spinor
takes the form $\Psi_{ABCD}=-4o_{(A}o_Bo_C\iota_{D)}$. Additionally
the Ricci spinor will be assumed to take the form
$\mu^2\alpha_{(A}\beta_{B)}\bar\alpha_{(A'}\bar\beta_{B')}$
where $\mu\in\R$ and $\alpha_A$ and $\beta_A$ are as in
(\ref{eq.nonnull}). The mixed invariants $N_1$ and $N_2$ are then given by
\begin{eqnarray}
N_1&=&4B(2AB+1)\label{eq.nnIIIn1}\\
N_2&=&4B^2\label{eq.nnIIIn2}
\end{eqnarray}
Following~\cite{p.r2} a necessary and sufficient condition for alignment
is that $Q_1=0$ where $Q_1$ is defined by
\begin{equation}
\label{eq.nnIq1}
Q_1=2I-4N_2+(N_1)^2
\end{equation}
In the case of Petrov type $III$ one has that $I=0$ and on
substituting (\ref{eq.nnIIIn1}) and (\ref{eq.nnIIIn2}) into (\ref{eq.nnIq1})
one obtains
\begin{equation}
\label{eq.nnIIIq1}
Q_1=64B^3A(AB+1)
\end{equation}
The {\sl non-repeatedly partially aligned} case is where $AB+1=0$
or $A=0$ but $B\neq 0$ and so $\iota_A$ is parallel to $\alpha_A$
or $\beta_A$ but $o_A$ is not parallel to either principal Ricci spinor.
Hence from (\ref{eq.nnIIIn2}), (\ref{eq.nnIq1}) and (\ref{eq.nnIIIq1})
one can see that {\sl necessary and sufficient
conditions for non-repeated
partial alignment} are $N_2\neq 0$ and $(N_1)^2=4N_2$.
In this case $N_1$ determines the pair $A,B$ (up to sign) and hence
the invariants $N_1,R_1$ form a complete set.

If $N_2=0$ then $N_1$ is also zero and $o_A$ is parallel
to $\alpha_A$. Thus, {\sl $N_2=0$ is a necessary and sufficient condition
for repeated partial alignment or total alignment}.  In these cases
$\Phi_{ABA'B'}=\phi_{AB}\bar{\phi}_{A'B'}$ where
 $\phi_{AB}=\mu(Ao_Ao_B+o_{(A}\iota_{B)})$ and there is no way
of determining $A$ from contractions of the Ricci and Weyl spinors.
A complete set of invariants is $\{A, R_1\}$ and
the partially and totally aligned cases can be separated using similar
methods to those used in the type $D$ case.

In the non-aligned case, $A$ and $B$ are determined by $N_1$ and $N_2$
and  so the Riemann spinor is completely fixed by $R_1,N_1$ and $N_2$.
The cross ratio $\chi$ of the four spin directions spanned respectively
by $o_A,\iota_A,\alpha_A$ and $\beta_A$ is given by equation
(\ref{eq.nnDcross}) and can be expressed in terms of $N_1$
and $N_2$  as
\begin{equation}
\label{eq.nnIIIcross}
\frac{(N_1)^2}{4N_2}=\left(\frac{1+\chi}{1-\chi}\right)^2
\end{equation}
The question of the significance of $N_1=0$, $N_2\neq 0$ arises.
In this case one can see from (\ref{eq.nnIIIcross}) that
$\chi=-1$ and hence the four principal spinors are {\sl harmonic}~\cite{p.r2}.
Equivalently, the flagpole directions of the four spin directions
form a square on the celestial sphere of null directions.
The other possibilities for the harmonic case
are $\chi=2$ or $\chi=\sfrac{1}{2}$ and these two values for $\chi$
are equivalent to $(N_1)^2=36N_2$. In the three harmonic cases
one has that $\alpha_A$ is parallel to
either $Ao_A+2\iota_A$, $2Ao_A+\iota_A$ and $Ao_A-\iota_A$ where
$\beta_A$ is as in (\ref{eq.nonnull}).

\paragraph*{Petrov type $II$}
The canonical form for the Petrov type $II$ Weyl spinor was  was given
in equation (\ref{eq.null2weyl}) and the dyad $o_A,\iota_A$
is fixed up to a finite group of transformations.
It will be assumed that the Ricci spinor takes the
same form as in the preceding discussion of the Petrov type $III$ case.

There are a number of possible types of alignment in the case
of a Petrov $II$ non-null Einstein-Maxwell field and the different types
will be listed before the interpretation of the mixed invariants
is discussed. Firstly, there is the {\sl non-aligned} case where the set
of five spinors consisting of the principal Weyl and Ricci spinors
all have different
directions in spin-space. In the {\sl non-repeatedly partially
aligned} case, one of the Ricci principal spinors aligns with
one of the non-repeated Weyl principal spinors, and there are no
other alignments, whereas in the {\sl non-repeatedly totally
aligned} case the pair of principal Ricci spinors aligns with
the pair of non-repeated Weyl spinors. If the repeated principal
Weyl spinor aligns with one of principal Ricci spinors and there
is no other alignment then the Weyl and Ricci spinors
are said to be {\sl repeatedly partially aligned}.
Finally the {\sl repeatedly totally aligned} case is where
the principal spinors of the Ricci spinor are both aligned with
principal Weyl spinors - one repeated
and one non-repeated.

In all cases, the Weyl spinor is completely determined
(up to sign) by the pure invariant $I$ and then the directions
of the Ricci principal spinors are determined by $A$ and $B$.
The expressions for $N_1$ and $N_2$ in terms of $A$ and $B$ are
\begin{eqnarray}
N_1&=&-2\frac{J}{I}\left(6B(BA^2+A+B)+1\right)\label{eq.nnIIn1}\\
N_2&=&\frac{I}{3}\left(6B(BA^2+A-2B)+2\right)\label{eq.nnIIn2}
\end{eqnarray}
In general this pair of equations  determine $A$ and $B$
in terms of $N_1, N_2, J$ and $I$. For example one has
\begin{equation}
\label{eq.nnIIB}
18B^2=1-\frac{3N_2}{I}-\frac{IN_1}{2J}
\end{equation}
and then the resulting solution for $B$ can be substituted back to find
$A$ (unless $B=0$). Noting that $I^3=6J^2$, it follows that
the set of invariants $I,N_1,N_2$ and $R_1$ determine the Ricci
and Weyl spinors, up to sign.

As in the type $III$ case, the necessary and sufficient condition
for any type of alignment is $Q_1=0$ where $Q_1$ is defined by (\ref{eq.nnIq1})
and in this case is given by
\begin{equation}
\label{eq.nnIIq1}
Q_1=24IB^2(A^2+1)\left((AB+1)^2+B^2\right)
\end{equation}
It can be confirmed that $Q_1=0$ is equivalent
to either $B=0$, $A=\pm i$ or $AB+1=\pm iB$, which correspond to the different
ways of aligning one of the principal Ricci spinors with one of the principal
Weyl spinors. In the repeatedly aligned cases (partially or totally) one
necessarily
has $B=0$ and from (\ref{eq.nnIIB}) one sees that {\sl a necessary
and sufficient condition for repeated partial or total alignment
is $2IJ=6JN_2+I^2N_1$}.  In
both the repeatedly aligned cases all real and complex mixed invariants
formed from contractions can be expressed in terms of the pure invariants
and a complete set of invariants is $\{I,R_1,A\}$.

If $B\neq 0$ (equivalently $6JN_2+I^2N_1\neq 2IJ$)
but $Q_1=0$ then the Ricci spinors must be non-repeatedly aligned -
either totally or partially. If one assumes total
alignment then it follows that $4J=IN_1$. Conversely if $4J=IN_1$ and $Q_1=0$
then it can be shown (from (\ref{eq.nnIq1}))
 that $6N_2=7I$ and equations (\ref{eq.nnIIn1})
and (\ref{eq.nnIIn2}) imply
\begin{eqnarray}
B(BA^2+A+B)&=&-\frac{1}{2}\label{eq.nnIIt1}\\
B(BA^2+A-2B)&=&\frac{1}{4}\label{eq.nnIIt2}
\end{eqnarray}
Solving the above gives $(A,B)=(\pm i,\pm\sfrac{i}{2})$ which is seen
to be equivalent to non-repeated total alignment. One therefore
has that {\sl necessary and sufficient conditions for non-repeated
total alignment are $N_1I=4J$ and $6N_2=7I$}. It also follows
that  {\sl  $Q_1=0$,
 $N_1I-4J\neq 0$, $N_1I+2J\neq 0$ are necessary and sufficient conditions
for non-repeated partial alignment} (the latter ensuring that $B\neq 0$).
The values of $A$ and $B$ can be determined from $N_1$ in the case of
partial alignment and so a complete set of invariants (ignoring sign
ambiguities) is $\{R_1, I,N_1\}$. In the totally aligned case only
pure invariants are required to form a complete set.

\paragraph*{Petrov type $I$}

In this case the Weyl spinor has four non-repeated principal spinors
and so there are three possibilities for alignment between
the Ricci and Weyl spinors : total, partial or non-aligned.
The canonical form for the Weyl spinor is given by (\ref{eq.Icanonical})
and the Ricci spinor will be assumed to take the same form as in the
preceding discussion of Petrov types $III$ and $II$. The invariants
$N_1$ and $N_2$ are given by the following rather lengthy expressions.
\begin{eqnarray}
N_1&=& -12\,\eta\chi A^{3}B^{2}+\left(12\left(\chi+1\right)\eta
\,B^{2}-18\eta\chi B\right)A^{2}\nonumber\\
& + & \left(-12\eta B^{2}+12
\left(\chi+1\right)\eta B-6\eta\chi\right)A\nonumber\\
& - & 6\eta B+2
\left(1+\chi\right)\eta\label{eq.nnIn1}\\
N_2&=&
9\eta^{2}\chi^{2}A^{4}B^{2}+\left(-\left(12\chi^{2}+12\chi
\right)\eta^{2}B^{2}+18\eta^{2}\chi^{2}B\right)A^{3}\nonumber \\
& + & \left(\left
(2+\chi+2\chi^{2}\right)6\eta^{2}B^{2}-18\left(1+
\chi\right)\chi\eta^{2}B+9\eta^{2}\chi^{2}\right)A^{2}\nonumber \\
& + & \left(
-12\left(1+\chi\right)\eta^{2}B^{2}+6\left(2+\chi+2\chi^{2}\right)
\eta^{2}B-6\left(\chi+1\right)\chi\eta^{2}\right)A \nonumber\\
&
 + & 9\eta^{2}B^{2}
 -6\left(\chi+1\right)\eta^{2}B -\left(\chi-
4\chi^{2}-4\right)\eta^{2}\label{eq.nnIn2}
\end{eqnarray}
{}From~\cite{p.r2} it follows that {\sl a necessary and sufficient
condition for at least one of the Ricci principal spinors
to align with one of the principal Weyl spinors is $Q_1=0$}
where in this case
 $Q_1$ factorises as follows
\begin{equation}
\label{eq.nnIq1f}
Q_1=144\,\eta^{2}AB\left(A-1\right)\left(\chi\,A-1\right)\left(AB+1
\right)\left(AB-B+1\right)\left(\chi\,AB-B+\chi\right)
\end{equation}
Noting that the principal Weyl spinors are $o_A$, $\iota_A$,
$o_A+\iota_A$ and $o_A+\chi\iota_A$, it is clear that the
vanishing of $Q_1$ is equivalent to at least partial alignment.
To determine a condition for total alignment, the easiest method is
to assume total alignment and examine the invariants. There are
actually nine different ways of aligning the principal Ricci spinors
with a pair of principal Weyl spinors, but the algebraically
simplest possibility would seem to be $A=B=0$. Assuming $A$ and
$B$ both zero the pure Weyl invariants are given by
(\ref{eq.null1i}) and (\ref{eq.null1j}) and the mixed invariants are given by
\begin{eqnarray}
N_1&=&2\eta(1+\chi)\label{eq.nnIalign1}\\
N_2&=&\eta^2(4\chi^2-\chi+4)\label{eq.nnIalign2}
\end{eqnarray}
Searching for relationships between $I,J,N_1$ and $N_2$ one finds
that $Q_2=0$ where
\begin{equation}
\label{eq.nnIq2}
Q_2=6N_1I-3(N_1)^3+8J
\end{equation}
Now it must be checked that $Q_1=Q_2=0$ is necessary and  sufficient for total
alignment of all types. To achieve this it is necessary to
assume partial alignment (i.e. $Q_1=0$) and, taking each
possible solution of $Q_1=0$ in turn,
factorise the resulting
expression for $Q_2$, thus solving the equation $Q_2=0$.
For example,
assuming $\beta_A$ and $\iota_A$ are parallel, that is $A=0$,  one has
\begin{equation}
\label{eq.nnIq2al}
Q_2=648\eta^3B(B-1)(B-\chi)
\end{equation}
Thus $Q_2=0$ is equivalent to $B=0,1$ or $\chi$ - the conditions
for $\alpha_A$ to align with one of the  other principal Weyl spinors
(note that $\eta\neq 0$ for Petrov type $I$). The other possible
cases can be dealt with similarly and it  may be verified
that {\sl $Q_1=Q_2=0$
is necessary and sufficient for total alignment}.

\section{Perfect Fluids}

For the purpose of this section it will be assumed that the
trace-free Ricci tensor $S_{ab}$
takes the form
\begin{equation}
\label{eq.pfricci}
S_{ab}=\frac{1}{4}(\mu+p)(4u_au_b-g_{ab})
\end{equation}
In the above, $\mu,p\in\R$, $u_a$ is a future pointing unit timelike
vector and the spacetime metric $g_{ab}$ has signature $-2$.
For each of the Petrov types, a canonical dyad $o_A,\iota_A$
is fixed by the Weyl spinor, up to some subgroup of $SL(2,\C)$, and determines
a corresponding complex null tetrad $l_a,n_a,m_a,\bar{m}_a$ in a
standard way~\cite[page 120]{p.r}. The complex null tetrad
satisfies
$l^an_a=-m^a\bar{m}_a=1$ with all other inner products
zero and thus $u_a$ can be expressed in terms of this tetrad
according to
\begin{equation}
\label{eq.pfu}
u_a=An_a+\left(\frac{1+2C\bar{C}}{2A}\right)l_a+Cm_a+\bar{C}\bar{m}_a
\end{equation}
for $C\in\C$, $A\in\R$, $A>0$. The components of $\Phi_{ABA'B'}$
in the corresponding dyad $o_A,\iota_A$ are then given below.
Bold-face indices are used to denote components
and the common factor $(\mu+p)$ has been omitted.
\begin{alignat}{3}
\Phi_{\bf 000'0'}&=\frac{1}{2}A^2 &\quad\Phi_{\bf 010'0'}&=-\frac{1}{2}AC &
\quad\Phi_{\bf 110'0'}&=\frac{1}{2}C^2\nonumber\\
& &\quad\Phi_{\bf 010'1'}&=\frac{1}{8}+\frac{1}{2}C\bar{C} &
\quad\Phi_{\bf 110'1'}&=-\frac{C(1+2C\bar{C})}{4A}\label{eq.pfric}\\
& & & &\quad \Phi_{\bf 111'1'}&=\frac{(1+2C\bar{C})^2}{8A^2}\nonumber
\end{alignat}
The Ricci scalar $R$ has the value $3p-\mu$ in this case
 and the other pure Ricci invariant
required is $R_1$ which haa
s the value $\sfrac{3}{16}(\mu+p)^2$ and is necessarily
non-zero. The Ricci spinor is uniquely
defined, up to sign, by
the unit timelike vector $u_a$ and the pure invariant $R_1$.
The role of the mixed invariants in this case is to determine $u_a$ in
terms of the canonical tetrad fixed by the Weyl tensor.
For perfect fluids, appropriate pure invariants are  $P_1$ and $P_2$
and defined by
\begin{eqnarray}
P_1&=&\frac{3M_3}{4R_1}\label{eq.pfp1}\\
P_2&=&\frac{3M_5}{16R_1}\label{eq.pfp2}
\end{eqnarray}
The corresponding tensor expressions for $P_1$ and $P_2$ are displayed below
\begin{eqnarray}
P_1&=&T_{abcd}u^au^bu^cu^d\label{eq.pfp1tens}\\
P_2&=&\sfrac{1}{16}\antiself{C}^{aghb}(C_{acdb}C_{gefh}+\dual{C}_{acdb}
\dual{C}_{gefh})u^cu^du^eu^f\label{eq.pfp2tens}
\end{eqnarray}
where $C_{abcd}$ is the Weyl tensor, $\dual{C}_{abcd}$
its dual and $\antiself{C}_{abcd}\equiv\sfrac{1}{2}(C_{abcd}+i\dual{C}_{abcd})$
is the {\sl anti-self dual} part.

It is of interest to discuss the behaviour of $P_1$ considered as a
quartic form on the space of unit timelike vectors i.e.
$P_1(t^a)=T_{abcd}t^at^bt^ct^d$. It is known~\cite{p.r}
that $P_1$ is necessarily positive but one can actually find
a lower bound for $P_1$ which varies with Petrov type. Before
proving a theorem concerning the lower bound of $P_1$, some notation
will be required. The {\sl electric} and {\sl magnetic} parts
of the Weyl tensor (with respect to some unit timelike vector $t_a$)
are denoted by $E_{ab}$ and $B_{ab}$ respectively and defined by
\begin{equation}
\label{eq.elmag}
E_{ab}=C_{acdb}t^ct^d\qquad B_{ab}=\dual{C}_{acdb}t^ct^d
\end{equation}
It should be noted that $E_{ab}$ and $B_{ab}$ are both symmetric and orthogonal
to $t_a$ and are thus elements of the vector space
$S=\{S_{ab}\vert S_{ab}=S_{ba}, S_{ab}t^b=0\}$. The vector space
$S$ is endowed with a negative-definite inner product defined by
total contraction. A {\sl duality rotation} of angle $\theta$ can be applied
to the Weyl tensor $C_{abcd}$ and transforms it to ${}^{(\theta)}C_{abcd}$
where
\begin{equation}
\label{eq.duality}
{}^{(\theta)}C_{abcd}=\cos(\theta) C_{abcd}+\sin(\theta) \dual{C}_{abcd}
\end{equation}
The following theorem concerning the lower bound of $P_1$ can now be
given.

\begin{theorem}
\label{th.bound}
If $P_1(t^a)=T_{abcd}t^at^bt^ct^d$ then it satisfies
the inequality
$4P_1\geq\vert I\vert$ for all possible unit timelike vectors $t^a$.
Furthermore, one has  $4P_1=\vert I\vert$ if and only if there
exists $\theta$ such that ${}^{(\theta)}C_{abcd}t^bt^c=0$.
\end{theorem}

\paragraph*{Proof}
Applying the Cauchy-Schwartz inequality to the inner product space $S$ defined
above one obtains
\begin{equation}
\label{eq.cauchy}
(E_{ab}B^{ab})^2\leq (E_{ab}E^{ab})(B_{ab}B^{ab})
\end{equation}
with equality if and only if $E_{ab}$ and $B_{ab}$ are linearly dependent.
Multiplying each side of (\ref{eq.cauchy}) by $4$ and adding
$(E_{ab}E^{ab})^2+(B_{ab}B^{ab})^2$   one obtains
\begin{equation}
\label{eq.ineq}
(E_{ab}E^{ab}-B_{ab}B^{ab})^2+4(E_{ab}B^{ab})^2\leq
(E_{ab}E^{ab}+B_{ab}B^{ab})^2
\end{equation}
again with equality if and only if $B_{ab}$ and $E_{ab}$ are parallel.
Using the definitions of $E_{ab}$ and $B_{ab}$ to expand the right hand side,
one finds that (\ref{eq.ineq}) is equivalent to
\begin{equation}
\label{eq.ineq2}
\left\vert(E_{ab}+iB_{ab})(E^{ab}+iB^{ab})\right\vert^2\leq
\left((C_{abcd}C^a{}_{ef}{}^d+
\dual{C}_{abcd}\dual{C}^a{}_{ef}{}^d)t^bt^ct^et^f\right)^2
\end{equation}
Finally it can be seen from~\cite{kramer} that the left hand side
of (\ref{eq.ineq2}) is equal to $\vert I\vert^2$ (note
that the definition of $I$ in that reference differs by a factor of
$1/2$ from the one used here) and using the definition
of $T_{abcd}$ in terms of the Weyl tensor~\cite{p.r}
it can be seen that the right hand side of (\ref{eq.ineq2})
is equal to $(4P_1)^2$. Consequently the inequality
\begin{equation}
\label{eq.ineq3}
\vert I\vert\leq 4P_1
\end{equation}
has been established. This inequality is an equality if and only
if $E_{ab}$ and $B_{ab}$ are linearly dependent, that is to say
if and only if there exists $\theta$ such that
$\cos(\theta)E_{ab}+\sin(\theta)B_{ab}=0$. This last equation
can be seen to be equivalent to ${}^{(\theta)}C_{abcd}t^bt^c=0$
and the proof of the theorem is complete.

Defining the complex symmetric tensor
$P_{ab}\equiv E_{ab}+iB_{ab}$ and noting that ${}^{(\theta)}C_{abcd}t^bt^c=0$
implies that $E_{ab}$ and $B_{ab}$ are linearly dependent,
it follows that $P_{ab}$ is a complex multiple of a real tensor
in this case. Hence $P_{ab}$ is diagonable with eigenvalues
of equal phase and therefore the Weyl tensor is of Petrov type $I$, $D$ or $O$.
In the Petrov type $I$ case one can see from~\cite{p.r2}
that the equality of the phases of the eigenvalues implies
that the cross-ratio of the four principal spinors
is real and hence the Weyl tensor belongs
to the subclass of Petrov type $I$ denoted by $I(M^+)$
as defined by Arianrhod and McIntosh~\cite{ar.mc}.
It also follows that, in the case of Petrov types $II$, $N$ and $III$,
the lower bound on $P_1$ cannot be attained. In fact
it will be seen from the calculations which follow
that $\vert I\vert=4P_1$ would imply that
the angle between $t_a$ and the (unique) repeated principal
null direction vanishes. The value of $4P_1$ therefore
approaches $\vert I\vert$ as $t_a$ approaches the repeated principal
null direction.

\paragraph*{Petrov type $N$}

In this case the freedom in the canonical dyad $o_A,\iota_A$ determined by
writing $\Psi_{ABCD}=o_Ao_Bo_Co_D$ may be used to make $C=0$
in (\ref{eq.pfu}). It can then be shown that $P_1=A^4$,
that is to say that $P_1=(u_al^a)^4$. The value of $A$
is therefore determined  by $P_1$ and then the Ricci
spinor is given by (\ref{eq.pfric}) and the pure invariant $R_1$.
A complete set of invariants for a  Petrov type $N$ perfect fluid
is therefore $\{R,R_1,P_1\}$.

\paragraph*{Petrov type $D$}

As in the previous two sections the canonical form $\Psi_{ABCD}=
6\eta o_{(A}o_B\iota_C\iota_{D)}$ will be used where the dyad
$o_A,\iota_A$ is defined up to reordering and the
rescaling $o_A\rightarrow\lambda o_A$
and $\iota_A\rightarrow\lambda^{-1}\iota_A$.
The corresponding complex null tetrad is therefore defined
up to arbitrary boosts in the $l-n$ plane and spatial rotations
in the $m-\bar{m}$ plane, which can be used to make $A=1$ and
$C=\bar{C}$.  In this tetrad the mixed invariant $P_1$ is given by
\begin{equation}
\label{eq.pfDp1}
P_1=\frac{\vert I\vert}{4}(1+12C^2+24C^4)
\end{equation}
The {\sl aligned} case is where $u_a$ lies in the $l-n$ plane and this
is equivalent to $C=0$, or $4P_1=\vert I\vert$ (the minimum
value of $P_1$). In all cases the
direction of $u_a$ is determined by $P_1$ via (\ref{eq.pfDp1}),
which determines $C$. The equation (\ref{eq.pfDp1}) appears
to give four possible values for $C$, but since $P_1>0$ \cite{p.r}
there is a unique real solution of (\ref{eq.pfDp1}) for
$C$ in terms of $P_1$ and $\vert I\vert$.
A complete set of invariants in this case is given by
(allowing for sign ambiguities)
the pure Ricci invariants $R$ and $R_1$, the pure
Weyl invariant $I$ and the mixed invariant $P_1$.

\paragraph*{Petrov type $III$}

As before the Petrov type $III$ Weyl spinor will be assumed
to take the canonical
form $-4o_{(A}o_Bo_C\iota_{D)}$ where the dyad $o_A,\iota_A$
is determined up to a sign change and hence determines a unique
complex null tetrad. If $u_a$ is given by (\ref{eq.pfu}) in this tetrad
then the mixed invariants $P_1$ and $P_2$ are expressed in terms of
$A$ and $C$ by the following equations.
\begin{eqnarray}
P_1&=&2A^2(1+8C\bar{C})\label{eq.pfIIIp1}\\
P_2&=&-2A^3\bar{C}\label{eq.pfIIIp2}
\end{eqnarray}
Two possible cases arise. The {\sl aligned} case is where $u_{[a}l_bn_{c]}=0$
and is characterised by $C=0$, or equivalently $P_2=0$.
The {\sl non-aligned} case corresponds to $P_2\neq 0$.
In all cases one can clearly determine $C$ and $A$ from $P_1$ and $P_2$.
Consequently a complete set of invariants for
the Riemann spinor in this case
is $\{R,R_1,P_1,P_2\}$.

\paragraph{Petrov type $II$}

Assume that the Weyl spinor takes the canonical form (\ref{eq.null2weyl})
where the principal spinors are $o_A$ (repeated) and
$o_A\pm i\iota_A$ (non-repeated) in the canonical dyad $o_A,\iota_A$.
There are essentially three cases  possible: the {\sl repeatedly
aligned} case where $u_a$ lies in the plane spanned by the repeated
principal null direction and one of the other principal
null directions; the {\sl non-repeatedly aligned} case where
$u_a$ lies in the plane spanned by the pair of non-repeated principal
null directions; and the {\sl non-aligned case}.

The expressions for the invariants $P_1$ and $P_2$ are
\begin{eqnarray}
P_1&=&\frac{3\eta\bar{\eta}}{2}\left[ 24A^4+24(C^2+\bar{C}^2)A^2
+(24C^2\bar{C}^2+12C\bar{C}+1)\right]\label{eq.pfIIp1}\\
P_2&=&\frac{3\eta^2\bar{\eta}}{8}\left[48A^4+24(2\bar{C}^2-C^2)A^2
-(24C^2\bar{C}^2+12C\bar{C}+1)\right]\label{eq.pfIIp2}
\end{eqnarray}
Recalling that $I=6\eta^2$ it can be seen
 that $4P_1$ tends to $\vert I\vert$ if $C=0$ and
$A$ tends to zero.
Notice also that
(\ref{eq.pfIIp1}) and (\ref{eq.pfIIp2}) are equations for
$C$ and $A$ in terms
of $I/J,P_1$ and $P_2$. Consideration of $Q_1\equiv4P_2+\eta P_1$
 (see (\ref{eq.pfIIq1}))  enables one to find an
expression for
$A^2$ in terms of $\bar{C}^2$ which can be substituted into
(\ref{eq.pfIIp2}) and the resulting equation split into real
and imaginary parts and solved to determine $C$. Thus a complete
set of invariants in this case is $\{I/J,R,R_1,P_1,P_2\}$.

The relationship between the invariants  and the various types
of alignment will now be examined. The repeated principal null direction
is spanned by the vector $l_a$ of the canonical tetrad and
the non-repeated principal null directions
are spanned by $p_a\equiv l_a+n_a-i(m_a-\bar{m}_a)$
and $q_a\equiv l_a+n_a+i(m_a-\bar{m}_a)$. The necessary and sufficient
condition for repeated alignment can be shown to be $A^2=-C^2$ (which
implies that $C$ is purely imaginary).
The full expression for $Q_1$ defined above is
\begin{equation}
\label{eq.pfIIq1}
Q_1=108\eta^2\bar{\eta}A^2(\bar{C}^2+A^2)
\end{equation}
and it is clear that $Q_1=0$ is equivalent to $A^2+C^2=0$ (cf~\cite{c.mcl})
and hence {\sl $Q_1=0$ is a necessary and sufficient condition
for repeated alignment}.
In the case of non-repeated alignment, one has that $1-2C^2=2A^2$
and $C+\bar{C}=0$ and under these assumptions
it is found that $Q_2\equiv \eta P_1-2P_2-\sfrac{9}{4}\eta^2\bar{\eta}=0$.
The expression for $Q_2$ in the general case is
\begin{equation}
\label{eq.pfIIq2}
Q_2=27\eta^2\bar{\eta}C(2C\bar{C}^2+2CA^2+\bar{C})
\end{equation}
It follows that $Q_2=0$ implies that $C$ is purely imaginary
 and either $2A^2=1-2C^2$ or $C=0$. The former possibility
implies non-repeated alignment, but to deal with the $C=0$
case one is required to consider $Q_3$ defined by
\begin{equation}
\label{eq.pfIIq3}
Q_3=7\eta^2(P_1)^2-64(P_2)^2+\frac{27\eta^2 M_4}{4R_1}
-\frac{189\eta^4\bar{\eta}^2}{4}
\end{equation}
If $C$ is purely imaginary and $2A^2=1-C^2$ then $Q_3=0$, but
if $C=0$ and $A$ is arbitrary then
\begin{equation}
\label{eq.pffIIq3c0}
Q_3=-2916\eta^4\bar{\eta}^2A^4(2A^2-1)(2A^2+1)
\end{equation}
Hence $Q_3=0$ is equivalent
(in the $C=0$ case) to $A=\pm1/\sqrt{2}$ which implies
$1-2C^2=2A^2$. It has thus been established that
{\sl necessary and sufficient conditions for non-repeated alignment
are $Q_2=0=Q_3$}.

\paragraph*{Petrov type $I$}

The actual forms of the invariants $P_1$ and $P_2$ in terms
of the components of $u_a$ are somewhat complicated and not too
illuminating but
a result can be proven concerning alignment of the Ricci tensor with the
Weyl tensor. The Ricci and Weyl tensors (and spinors) will
be said to be {\sl aligned} when the unit timelike
Ricci eigenvalue is parallel to the timelike member of the
canonical orthonormal Petrov tetrad~\cite{e.k}.

It was mentioned earlier in this section that $P_1$ can be considered
as a quartic form on unit timelike vectors given by
$P_1(t^a)=T_{abcd}t^at^bt^ct^d$. Now define the quadratic form
$\gamma(t^a)\equiv g_{ab}t^at^b$ and denote the derivatives
of $\gamma$ and $P_1$ at a point $t^a$  in the tangent space by $D_t\gamma$ and
$D_tP_1$ respectively. It follows from the theory of Lagrange
multipliers~\cite[page 211]{abraham} that $t^a$ is a critical point
of $P_1$ restricted
to $\gamma^{-1}(1)$ if and only if there exists $\lambda\in\R$
such that $D_tP_1=\lambda D_t\gamma$. It can then be seen that this condition
is equivalent to
\begin{equation}
\label{eq.p1precrit}
4T_{abcd}t^bt^ct^d=2\lambda t_a
\end{equation}
or
\begin{equation}
\label{eq.p1crit}
t_{[a}T_{b]cde}t^ct^dt^e=0
\end{equation}

It will now be shown that (\ref{eq.p1crit}) is satisfied with
$t_a$ replaced by $u_a$ if and only if the Ricci and Weyl tensors
are aligned. That is to say, it will be shown that (\ref{eq.p1crit})
is equivalent to $t_a$ being the timelike member of the canonical
Petrov tetrad (which is unique up to sign).
 If one defines
the {\sl self-dual} part of the Weyl tensor $\self{C}_{abcd}$
as $\sfrac{1}{2}(C_{abcd}-iC_{abcd})$, so that the anti self-dual part
of the Weyl tensor ($\antiself{C}_{abcd}$) is the conjugate of the
self-dual part, then $T_{abcd}$ is given by~\cite{p.r}
\begin{equation}
\label{eq.p1self}
T_{abcd}=\self{C}^p{}_{ab}{}^q\antiself{C}_{pcdq}
\end{equation}
Now suppose that $t_a,x_a,y_a,z_a$ is an orthonormal tetrad
and define $F_{ab}=2t_{[a}x_{b]}$, $G_{ab}=2t_{[a}y_{b]}$ and
$H_{ab}=2t_{[a}z_{b]}$. If one then defines
$\self{F}_{ab}=\sfrac{1}{2}(F_{ab}-i\dual{F}_{ab})$,
$\self{G}_{ab}=\sfrac{1}{2}(G_{ab}-i\dual{G}_{ab})$ and
$\self{H}_{ab}=\sfrac{1}{2}(H_{ab}-i\dual{H}_{ab})$ the self-dual part of the
Weyl tensor can be expanded as follows
\begin{eqnarray}
\self{C}_{abcd}&=&\sigma_1\self{F}_{ab}\self{F}_{cd}+
     \mu_1(\self{G}_{ab}\self{H}_{cd}+\self{H}_{ab}\self{G}_{cd})\nonumber\\
&+&\sigma_2\self{G}_{ab}\self{G}_{cd}+
     \mu_2(\self{H}_{ab}\self{F}_{cd}+\self{F}_{ab}\self{H}_{cd})
     \label{eq.selfweyl}\\
&+&\sigma_3\self{H}_{ab}\self{H}_{cd}+
     \mu_3(\self{F}_{ab}\self{G}_{cd}+\self{G}_{ab}\self{F}_{cd})\nonumber
\end{eqnarray}
In the canonical Petrov tetrad  one has that $\mu_1=\mu_2=\mu_3=0$
but in a general tetrad $\sigma_i$ and $\mu_i$ are arbitrary complex
numbers, only restricted by $\sigma_1+\sigma_2+\sigma_3=0$.
Now recall that $x_a, y_a$ and $z_a$ are arbitrary for a fixed $t_a$
(subject to orthogonality relations) and so it may be assumed
that they are chosen so that $B_{ab}$ defined by
(\ref{eq.elmag}) assumes diagonal form. The consequences of
this restriction are that $\mu_1, \mu_2$ and $\mu_3$ are real and
that the eigenvalues of $B_{ab}$ are $-(1/4)$ times
the imaginary parts of $\sigma_1$, $\sigma_2$ and $\sigma_3$.
If the eigenvalues of $B_{ab}$ are distinct then the tetrad
$t_a,x_a,y_a,z_a$ is now fixed. If there is one pair of equal eigenvalues
then it may be assumed, without loss of generality,
that $\Im(\sigma_1-\sigma_2)=0$
 and
 the eigenvectors with equal eigenvalues are $x_a$ and $y_a$ and
then the further freedom in the tetrad can be used to set
$\mu_3=0$. The tracefree nature of $B_{ab}$ means that
any further eigenvalue degeneracy would force $B_{ab}=0$
and so in this case one
can diagonalise $E_{ab}$ and then set $\mu_1=\mu_2=\mu_3=0$.

Using (\ref{eq.p1self}) and (\ref{eq.selfweyl}) the condition
(\ref{eq.p1crit}) can be shown to be equivalent to the following three
equations
\begin{eqnarray}
\mu_3\cdot\Im\left(\sigma_1-\sigma_2\right)&=&
0\label{eq.crit1}\\
\mu_2\cdot\Im\left(\sigma_3-\sigma_1\right)&=&
0\label{eq.crit2}\\
\mu_1\cdot\Im\left(\sigma_2-\sigma_3\right)&=&
0\label{eq.crit3}
\end{eqnarray}
Assuming
that $B_{ab}$ has no eigenvalue degeneracies the above
clearly implies that all the $\mu_i$ vanish.
If there is an eigenvalue degeneracy of the form discussed
earlier, but $B_{ab}\neq 0$,
 then it may be assumed that
 $\mu_3=0$ and  then (\ref{eq.crit2}) and (\ref{eq.crit3})
imply $\mu_2=\mu_1=0$. Finally, if $B_{ab}=0$ then all the $\mu_i$
automatically vanish. Conversely if all the $\mu_i$ vanish
then equations (\ref{eq.crit1})-(\ref{eq.crit3}) are identically satisfied
and hence (\ref{eq.p1crit}) is satisfied.
It has therefore been shown that the critical points of $P_1(t^a)$
restricted to $\gamma^{-1}(1)$ are merely the two possible
timelike members of the canonical Petrov tetrad, which differ only
by a sign. Since $P_1(t^a)=P_1(-t^a)$
it may be assumed that $P_1$ is further restricted to future pointing members
of $\gamma^{-1}(1)$ and it has a unique critical point on this set.
It will now be shown that this critical point is the global minimum.
Since $\mu_i=0$ in the canonical Petrov tetrad (which will be denoted by
$v_a,x_a,y_a,z_a$) one can expand $\self{C}_{abcd}$ in this tetrad as
(writing $\sigma_i=\alpha_i+i\beta_i$)
\begin{eqnarray}
\self{C}_{abcd}&=&(\alpha_1\self{F}_{ab}\self{F}_{cd}+\alpha_2\self{G}_{ab}
\self{G}_{cd}+\alpha_3\self{H}_{ab}\self{H}_{cd})\nonumber\\
&+&i(\beta_1\self{F}_{ab}\self{F}_{cd}+\beta_2\self{G}_{ab}
\self{G}_{cd}+\beta_3\self{H}_{ab}\self{H}_{cd})\label{eq.weyldiag}
\end{eqnarray}
where the self-dual bivectors
have been  redefined in terms of  $F_{ab}=2v_{[a}x_{b]}$,
$G_{ab}=2v_{[a}y_{b]}$ and
$H_{ab}=2v_{[a}z_{b]}$.
The decomposition of the self-dual Weyl tensor into the two bracketed
quantities above gives rise to a corresponding decomposition
of the Weyl tensor $C_{abcd}=\uone{C}_{abcd}+\utwo{C}_{abcd}$.
It may then be verified that ${}^{(\pi/2)}\uone{C}_{abcd}v^bv^c=0$
and $\utwo{C}_{abcd}v^bv^c=0$ where ${}^{(\theta)}C_{abcd}$ was defined
by equation~(\ref{eq.duality}). The minimum values of the quartic
forms associated with $\uone{T}_{abcd}$ and $\utwo{T}_{abcd}$ are
shown  by theorem~\ref{th.bound}
 to occur at $v^a$. Since the following can be shown
to hold for any timelike vector $t^a$
\begin{equation}
\label{eq.sum}
T_{abcd}t^at^bt^ct^d=\uone{T}_{abcd}t^at^bt^ct^d+\utwo{T}_{abcd}t^at^bt^ct^d
\end{equation}
$P_1(t^a)$ has a minimum at $v^a$ also and the minimum
value is given by
\begin{equation}
\label{eq.minimum}
P_1(v^a)=\frac{1}{16}\sum_{i=1}^3\, {\vert\sigma_i\vert^2}
\end{equation}

The following theorem has therefore been established
\begin{theorem}
If the Weyl tensor is of Petrov type $I$ and $u^a$
is a future pointing unit timelike vector then the following are equivalent.
\begin{description}
\label{th.I}
\item{(i)} $u_{[a}T_{b]cde}u^cu^du^e=0$
\item{(ii)} $u^a$ is the timelike member of the canonical
Petrov tetrad.
\item{(iii)} $u^a$ is the unique global minimum of
the quartic form $t^a\mapsto T_{abcd}t^at^bt^ct^d$,
restricted to future pointing unit timelike vectors. The minimum
value is given by~(\ref{eq.minimum}).
\end{description}
\end{theorem}
It is remarked that the condition $u_{[a}T_{b]cde}u^cu^du^e=0$ can
be expressed in terms of a perfect fluid Ricci tensor to give
a necessary and sufficient condition for alignment.

\section*{Acknowledgements}
The author acknowledges the financial support of the European
Union in the form of a Human Capital and Mobility post-doctoral
fellowship.

\end{document}